\documentclass[12pt]{article}
\usepackage{amsfonts}
\usepackage{amsmath}
\usepackage{amssymb}
\usepackage{graphicx}

\begin{document}

{\centering{\Large \textbf{Quantum Correction for Newton's Law of Motion}}
\bigskip }

{\large T.F. Kamalov} \bigskip

\textit{Theoretical Physics Department, Moscow Institute of Physics and
Technology\newline
Institute str., 9, Dolgoprudny, Moscow Region, 141700, Russia\newline
}{\small E-mail: t.kamalov@phystech.edu} \smallskip

\abstract{A description of the motion in noninertial reference frames by means of the inclusion of high time derivatives is studied. Incompleteness of the description of physical reality is a problem of any theory, both in quantum mechanics and classical physics. The ``stability principle'' is put forward. We~also provide macroscopic examples of noninertial mechanics and verify the use of high-order derivatives as nonlocal hidden variables on the basis of the equivalence principle when acceleration is equal to the gravitational field. Acceleration in this case is a function of  high derivatives with respect to time. The~definition of dark metrics for matter and energy is presented to replace the standard notions of dark matter and dark energy. In the Conclusion section, problem symmetry is noted for noninertial mechanics.}

\section{Introduction}

The problem of physics axiomatizing being one of Hilbert's problems entails a
search for a unified axiom of both classical and quantum physics. In this
paper, the incompleteness problem of the quantum-mechanical description of 
physical reality is replaced with the incompleteness problem   of
classical physics. The implementation of the search for a unified axiom
of classical and quantum physics is also suggested through complementing
classical physics. This is related to quantum physics
being much richer in variables than classical physics, and complementing classical
physics with hidden variables is more reasonable than doing it with quantum
physics, which has been practiced for the last 100 years by numerous authors
in the effort to sew classical and quantum physics together.

\section{Why Newton's Law of Motion is a Second-Order Derivative Equation}

Since 1935, the contradiction of classical and quantum mechanics, and the
search for a satisfactory quantum axiom have been important
problems, but the search for this quantum axiom has been unsuccessful. A more
successful solution in providing the consistency of classical and quantum
physics would be a unified axiom. According to Gödel's theorem,
there are provisions in any theory  that cannot be proven in the framework of
this theory, and that no theory is complete. The axioms of any
theory have not been proved, but guessed, so any system of axioms can be
replaced by another. The main idea of Newton's laws in \textit{{Principia}} 
postulates a
description dynamics of mechanical systems with second-order differential
equations. There are cases of describing reality using higher-order
differential equations, but this is not Newtonian
mechanics. It is difficult to find an inertial reference frame
since there always exist random weak external fields and forces, but we can
assume that an inertial reference frame theoretically exists. A possible
example of non-Newtonian mechanics is quantum mechanics. A noninertial
reference frame is needed in order to add one of the most important
properties of micro-objects of quantum mechanics---nonlocality. In this
case, the role of nonlocal hidden variables is played by acceleration and
its higher derivatives with respect to time. In a noninertial reference
frame, the oscillations of two classical particles  correlate since the
acceleration and its high-order derivatives do not depend on their
coordinates. The description of mechanical systems in noninertial mechanics
is performed using high-order derivatives of differential equations. Let us assume that $%
q$ are  coordinates of a noninertial reference frame. Then, averaged $q
$ is denoted by $Q$:

\begin{center}
$Q=\left\langle q(t)\right\rangle =\int\limits_{-\tau }^{+\tau }\psi ^{\ast
}q\psi dt$.
\end{center}

Here, $\tau $ is a time interval for averaging, and

\begin{equation}
\psi =\psi _{0}e^{i\frac{\Delta p\Delta q+\Delta E\Delta t}{\hbar }}=\psi
_{0}e^{i\frac{\frac{\Delta p}{\Delta t}-F}{\frac{\hbar }{\Delta q\Delta t}}%
}=\psi _{0}e^{i\frac{f_{0}}{f_{Q}}}
\end{equation}
is wave function with  inertial force $f_{0}$ dependent on high-order
derivative coordinates on time; $f_{Q}$~corresponds with inertial forces
and constant force.

Noninertial reference frames are a method for describing the influence of
random fields on both the particle to be described and the observer. The
transition from a noninertial reference frame to an inertial one causes a
free particle to randomly oscillate, correlating with vibrations of other
free particles. Transformations of noninertial reference frames differ from
Galileo--Lorentz transformations by residual terms in the Taylor expansion.
Then, free particles in inertial reference frames are described with
uncertainty in coordinates and momentum, time, and energy, equal to the rest
of the terms of the Taylor expansion. If the transformation of a
noninertial reference frame to another, described by the Taylor expansion,
contains a remainder term with index N, then we can say that this free
particle conserves its time derivative of the N-th order. Such a free
particle is described by N derivatives and preserves this state until
interactions with other bodies (forces) perturb this state. If such a
particle interacts with other bodies (a~force acts on it), then the dynamics
of such a particle is described by differential equations of the \mbox{(N + 1)-th}
order. In other words, the influence of force adds one more derivative to
the description of particle dynamics. Considering a particle in an inertial
reference frame instead of noninertial ones, one  either introduces
inertial forces, that is, a change from higher-derivative description to the
description without higher derivatives, but with inertia forces, or takes
into account remainder terms of Taylor~expansion.

Modern physics (both classical and quantum) is the physics of inertial
reference frames. The case of a noninertial reference frame usually comes
down to the introduction of inertia forces into the inertial reference frame. The use of inertia forces makes it possible to reduce the problems of
the dynamics of a physical system in a noninertial reference frame to the
tasks in an inertial reference frame by artificially introducing inertial
forces or applying the d'Alembert principle. At the same time, an inertial
reference frame does not exist in nature, since any reference frame is
always influenced by infinitesimal disturbing fields or forces. In this
study, we propose to consider only noninertial reference frames as real.
Since the introduction of the d'Alembert principle to the present, the
reality of inertia forces has been a debated issue. The question of the reality of inertia forces can be reduced to the question
of the reality of inertial reference frames. How can one describe physical
systems in noninertial reference frames without introducing inertia
forces? The reference frame is  inertial if Newton's laws hold. 
They postulate the description of physical systems by
second-order differential equations. The rejection of higher-order
derivatives of coordinates is associated with the problem of inertia forces
in inertial reference frames. Thus, to answer the above question, we must consider a more general case of higher-order differential
equations and expand classical physics with a description using higher-order
derivatives. The transition from an inertial reference frame to a
noninertial one without introducing inertia forces means a transition from
a description of physical systems of second-order differential equations to
their description using higher-order differential equations. The rejection
of the use of higher-order derivatives with respect to coordinates in
classical Newtonian physics does not mean that they do not exist; they exist
in certain cases, but this is not Newtonian physics.

In the most general case, a transformation from the noninertial reference frame to another  can~be~expressed~as:

\begin{center}
\bigskip $Q=q(\tau ,\overset{\cdot }{q},\overset{\cdot \cdot }{q},\overset{%
\cdot \cdot \cdot }{q},...,q^{(n)})$

$T=t$.
\end{center}

Conversion of coordinates of a point particle between two noninertial
reference frames, provided that $\tau $ is a time interval for averaging, is
expressed as

\begin{center}
$Q=q(t)+\overset{\cdot }{q}(t)\tau +\Delta q(t),$

$\Delta q(t)=\sum_{k=2}^{n}(-1)^{k}\frac{1}{k!}\tau ^{k}q^{(n)}(t)$.
\end{center}

The same holds for momentum:

\begin{center}
\bigskip $P=p(t)+\overset{\cdot }{p}(t)\tau +\Delta p(t),$

$\Delta p(t)=\sum_{k=2}^{n}(-1)^{k}\frac{1}{k!}\tau ^{k}p^{(n)}(t)$
\end{center}

\bigskip

\begin{equation}
\left\langle P\right\rangle =\frac{1}{2}[p(t+\tau )+p(t-\tau )].
\end{equation}

Here, $\Delta q(t)$ and $\Delta p(t)$ are remainder terms of the Taylor
expansion. Remainder terms $\Delta q(t)$ and $\Delta p(t)$ in a noninertial
reference frame may be interpreted as coordinate and
momentum uncertainties of a point particle in this reference system. In quantum mechanics,
  coordinate and momentum uncertainties of a microparticle obey  
rule 
\begin{equation}
\Delta q(t)\Delta p(t)\geq \hbar /2.
\end{equation}

In noninertial physics a general uncertainty relation can be introduced ,
as there always exist random small fields and forces influencing either the
very system to be described or an observer, that is,

\begin{center}
$[\sum_{k=2}^{n}(-1)^{k}\frac{1}{k!}\tau
^{k}x^{(k)}(t)][\sum_{k=2}^{n}(-1)^{k}\frac{1}{k!}\tau ^{k}p^{(k)}(t)]\leq
K/2$
\end{center}

The supremum of the difference of the action function in noninertial
reference frames (with higher time derivatives of the generalized
coordinate) from the classical mechanics action functions (without higher
derivatives) is that, in this case, higher derivatives are nonlocal additional
variables that disclose the sense of the classical analogue $K$ of Planck's
constant.  $K$ constant defines the supremum of the influence of random
fields onto the physical system and the observer. We  analyze this case
in terms of a noninertial reference frame. In this case, $K$ defines the
supremum of the difference between a noninertial and an inertial reference frame:

\begin{equation}
\sup \left\vert S(q,\overset{\cdot }{q},\overset{\cdot \cdot }{q},\overset{%
\cdot \cdot \cdot }{q},...,q^{(n)},...)-S(q,\overset{\cdot }{q})\right\vert
=K.
\end{equation}

Action functions in higher derivatives of a noninertial reference frame
describe physical systems dynamics and differ from the action function
neglecting random fields, which are accounted for via noninertial reference frame. In our case, a classical space is featured by an infinite number of
variables, same as one of Hilbert's. In the search for a unified axiomatic,
classical constant $K$  coincides with the quantum constant, i.e., Planck's
constant $\hslash $. In this approach, the estimate of the Planck constant
may be determined by higher derivatives playing the role of nonlocal
hidden variables.

In this case the state of quantum object, we can  describe

\begin{center}
$\left\vert \psi (t)\right\rangle =\left\vert q,\overset{\cdot }{q},\overset{%
\cdot \cdot }{q},\overset{\cdot \cdot \cdot }{q},...,q^{(n)}\right\rangle
=\left\vert Q(t)\right\rangle .$
\end{center}

The transfer object from point $1$ to point $2$ is

\begin{center}
$\left\langle Q_{1},t_{1}|Q_{2},t_{2}\right\rangle
=\int_{t_{1}}^{t_{2}}DQ\exp (\frac{i}{\hbar }L(Q))dt.$
\end{center}

The introduced function can be represented as

\begin{center}
$A_{Q(R)}=\left\langle q(t),\overset{\cdot }{q}(t),\overset{\cdot \cdot }{q}%
(t),...,q^{(n)}(t)|[i\hbar \frac{\partial }{\partial t}-H]|q(t),\overset{%
\cdot }{q}(t),\overset{\cdot \cdot }{q}(t),...,q^{(n)}(t)\right\rangle dt$.
\end{center}

\section{Stability Principle}

Classical mechanics describes a stable trajectory and noninertial
mechanics to add instability to random trajectories with high-order-derivative
variables. The stability condition in calculations of mechanical
trajectories was put forward in publications by Chetayev \cite{2}. According
to him, \textquotedblleft stability is probably an essentially general
phenomenon that has to manifest itself in the principal laws of
Nature.\textquotedblright\ In his opinion, stability is not a mere
casualness, but rather a consequence of a system being affected by
persistent infinitesimal perturbations that, no matter how small, affect
the state of a mechanical system. The condition of stability usually used in
mechanics can be extended to other areas of physics. In this case, the
condition of stability can be named the stability principle. The stability
principle is a generalization of basic fundamental physical laws, such as
the least-action principle, Newton's laws, Euler--Lagrange equations, and Schrödinger's
 equation. Our definition of the
stability condition was extended to  other areas of physics. Let us
define a stable state of a physical system through the stability principle.

{Stability principle:} {State }$A$ {of a physical
system is considered stable if it returns to the initial state after
finishing the action of external factors, and the variance of the variable
with itself is zero } $Var(A)=\sigma _{A}=0$.

We considered noninertial reference frames due the influence of the
background of random fields and waves because the variance of the action
function for an unstable trajectory with itself can be represented as $%
Var(S_{r})=\sigma _{Sr}=K$, and the complex variance can be defined in the form

\begin{center}
$Var(S_{c})=Var(S(q,\overset{\cdot }{q}))+iVar(S_{r}(q,\overset{.}{q},%
\overset{\cdot \cdot }{q},\overset{\cdot \cdot \cdot }{q},...,q^{(n)},...)=%
\sigma _{C}=\sigma _{S}+i\sigma _{Sr}$.
\end{center}
where variance for the classical stable trajectory  is $\sigma _{S}=0$;
 for an unstable trajectory, it is $\sigma _{Sr}=K$.

\section{Quantum Correlations and Illusion of Superluminal Interaction}

By discussing the nonlocality of entangled-state quantum correlations for
observers Alice and Bob, we see: The emerging illusion
of transfer from $A$ to $B$, or the interaction of entangled quantum objects in $%
A$ and $B$ follows from experimentally observed correlation of their states.
So, it would be correct to  not only negate faster-than-light interaction or
transfer, but the very fact of any interaction or transfer. The~existence of
quantum correlations and the nonlocality of micro-object quantum states may be
described by the noninertial nature of a noninertial reference. In other words,
the existence of quantum nonlocality and quantum correlations means an illusion
rather than the realness of any transfer or faster-than-light interaction of
these objects.

Let us perform an imaginary experiment of the classical analogue of
teleportation of quantum polarization states of biphotons.

For this purpose, let us consider the classical analogue of teleportation of
biphoton-polarization-state quantum entanglement. A classical analogue of
this situation may be considered on the example of newspapers with news
printed in  city $O$ and sent to cities $A$ and $B$.

If a reader in  city $A$ reads the news, then the coincidence of their
information with that in $B$ may be described with a nonzero correlation
factor. This is so because  news information in A and B  correlates
with a nonzero factor.

The complete match of  news information could only
occur provided that readers $A$ and $B$ read newspapers with the same title and
of the same date.

If the newspapers are different but both of the same date, then the
correlation factor is not unity, but at the same time, it is not
zero. To achieve the complete match of  news information with  correlation-factor unity,  reader $A$  advises  reader $B$ on both the title
 and  date of the newspaper.

To provide teleportation of biphoton quantum states from $A$ to $B$, we may
consider a primary photon that, with the aid of a nonlinear crystal (e.g., 
$BBO$), is split into two photons in  point $O$  with vertical $H$ and
horizontal $V$ polarizations. Photon $B$ may be compared with photon $C$,
entangled with photon $D$. Therefore, in points $A$ and $D$, measurements of
polarizations of the photons  always coincide.

Let us repeat the proof of  Bell's theorem incorporating influences of
any random fields, waves, or~forces onto both particles $A$ and $B$, and the
observers. We~consider here a noninertial reference frame. We may consider
that, in the inertial reference frame, these particles are influenced by
random inertia forces that, due to the equivalence principle, can be
described by random metrics.

\section{Quantum Correction to Newton's Second Law}

Ostrogradsky formalism \cite{1} using a Lagrange function is

\begin{center}
$L=L(q,\dot{q},\ddot{q},...,q^{(n)},...)$,
\end{center}

but not

\begin{center}
$L=L(q,\dot{q})$.
\end{center}

The Euler--Lagrange equation in this case follows from the least-action principle \cite{3,4,5,6}:

\begin{center}
$\delta S=\delta \int L(q,\overset{\cdot }{q},\overset{\cdot \cdot }{q},%
\overset{\cdot \cdot \cdot }{q},...,q^{(n)})dt=\int \sum_{n=0}^{N}(-1)^{n}%
\frac{d^{n}}{dt^{n}}(\frac{\partial L}{\partial q^{(n)}})\delta q^{(n)}dt=0$.
\end{center}

Alternatively,

\begin{center}
$\frac{\partial L}{\partial q}-\frac{d}{dt}\frac{\partial L}{\partial 
\overset{\cdot }{q}}+\frac{d^{2}}{dt^{2}}\frac{\partial L}{\partial \overset{%
\cdot \cdot }{q}}-\frac{d^{3}}{dt^{3}}\frac{\partial L}{\partial \overset{%
\cdot \cdot \cdot }{q}}+...+(-1)^{n}\frac{d^{n}}{dt^{n}}\frac{\partial L}{%
\partial q^{(n)}}+...=0$.
\end{center}

This equation can be written in the form of a corrected Newton's second law of
motion in noninertial reference frames:

\begin{center}
$F-ma+f_{0}=0$.
\end{center}

Here,

\begin{center}
$f_{0}=mw=w(t)+\overset{\cdot }{w}(t)\tau +\sum_{k=2}^{n}(-1)^{k}\frac{1}{k!}%
\tau ^{k}w^{(n)}(t)$
\end{center}
is a random inertial force (1) that can be represented by Taylor expansion
with high-order derivatives coordinates on time

\begin{center}
$F-ma+\tau m\overset{\cdot }{a}-\frac{1}{2}\tau ^{2}ma^{(2)}+...+\frac{1}{n!}%
(-1)^{n}\tau ^{n}ma^{(n)}+...=0$
\end{center}
in  inertial reference frame $w=0$.

\section{Dark Metric for Matter and Energy}

From \cite{7}, it follows that the phase space of coordinates and high-order
derivatives gives the corrected Newton's formula for gravitational potential

\begin{center}
$\varphi =GM\exp s/r$,
\end{center}
where $\varphi $, potential; $G$, gravitational constant; $s=-2GM$,
constant; and $M$, mass.

On the one hand,  force $F$ is expressed using infinite Taylor
expansion. On the other hand,  gravitational force $F_{g}$ can also be
represented as a series, as follows from the principle of equivalence. If
this series is replaced by an exponential \cite{7}, then we can write metric

\begin{center}
$ds^{2}=\exp (-\frac{r_{0}}{r})dt^{2}-\exp (\frac{r_{0}}{r}%
)dr^{2}-r^{2}d\theta ^{2}-r^{2}\sin ^{2}\theta d\phi ^{2}$,
\end{center}
which we call the dark metric \cite{8}, where $r_{0}=2GM$.

The dark metric is the asymptotic of the Schwarzschild metric for $r_{0}<r$
\cite{9,10}. The definition of dark metrics for matter and energy is presented to replace the standard notions of dark matter and dark energy. The dark metric can also be obtained from the standard metric:

\begin{center}
$ds^{2}=B(r)dt^{2}-A(r)dr^{2}-r^{2}d\theta ^{2}-r^{2}\sin ^{2}\theta d\phi
^{2}$.
\end{center}

Conditions $A(r)B(r)=1$ and $\lim A(r)=B(r)=1$ for $r\rightarrow \infty $\
must be satisfied for the standard metric. The dark metric also satisfies to these conditions. Gravitational forces are presented as a series with changing signs.

\section{Macroexamples of Noninertial Mechanics}

The behavior of macroscopic mechanical systems in noninertial reference frames can be described by higher-order differential equations. Here, we
consider the case when the contribution of higher derivatives is small
compared to lower ones. Therefore, at this stage, we restricted ourselves to
only the third derivatives of the coordinates with respect to time. There
are many examples of the description of mechanical systems in noninertial
reference frames \cite{3,4,5,6} due to the influence of the backgrounds of random
fields and waves. Theoretical descriptions of such cases do not always fully
describe the physical reality of  processes occurring in this process.
Such cases include Kapica's pendulum, the movement of bulk materials
upwards, against the action of gravity, and Chalomey's pendulum \cite{11}. For describing vibrating mechanical systems, the principle of least
action is traditionally used to obtain critical states of mechanical
systems. All such cases are described by second-order differential
equations. In this case, the direction of the resultant force remains
uncertain. This is the main disadvantage of this method of description.
Using the extended Newton's second law \cite{9}

\begin{equation}
F-ma+\tau m\overset{\cdot }{a}-\frac{1}{2}\tau ^{2}ma^{(2)}+...+\frac{1}{n!}%
(-1)^{n}\tau ^{n}ma^{(n)}+...=0,
\end{equation}
where $\tau =1/\omega $ is the averaging time during the transition from the
micro- to the macroworld, which is inverse to the average cyclic frequency, we
obtain the direction of the resultant force that coincides with the
direction of the motion. In \cite{9}, the behavior of such systems is described
by introducing experimental vibration forces. The introduction of vibration
forces in these cases is not justified and is axiomatically~introduced.

Here, we  use a third-order differential equation. This allows to first
obtain the correct direction of the resultant force. Second, it explains its
occurrence and does not contradict  already known descriptions.

Comparing the two descriptions: the differential equations of the second
 and  third order can  argue the consistency of these two
descriptions. Indeed, in mathematics, there is a method of transition from
higher-order differential equations to lower ones by changing variables. In
our case, from a third-order differential equation, we can go to two
equations of an order not higher than the second.

For example, consider the description of  Kapica's pendulum using 
differential Equation (6), limiting ourselves to the third order of the
derivative of the coordinate with respect to time%
\begin{equation}
F-ma+\tau m\overset{\cdot }{a}=0.
\end{equation}

Or
\begin{center}
$\qquad \qquad \qquad F=ma-mj\tau, \qquad \qquad $
\end{center}
where $j=\overset{\cdot }{a}=\frac{d^{3}q}{dt^{3}}$ is third-order
derivative coordinate $q$ on the time named Jerk\, and$\ \tau =1/\omega $ is
the averaging time during the transition from the micro- to the macroworld,
the opposite of  average cyclic~frequency. \qquad

Using the substitution, we get%
\begin{equation}
F+V=ma,
\end{equation}
where  vibration force $V$ is equal to%
\begin{equation}
V=mA\omega ^{2}\sin \omega t.
\end{equation}

Thus, we  showed that Equation (6) can be replaced by two others, Equations (7)
and (8). In this case, the description with high-order derivatives of
mechanical systems is more complete than the description with second-order
derivatives \cite{12,13,14,15,16,17,18,19,20,21,22,23}.

\section{Verifications of High-Order Derivatives as Nonlocal Hidden
Variables}

The role of high-order derivatives as hidden variables can be verified by
using the equivalence principle when acceleration is equal to the gravitational
field. Then, the correlation factor for entangled photons polarization
measurements may be presented as

\begin{equation}
\left\vert M\right\vert =\left\vert \left\langle AB\right\rangle \right\vert
=\left\vert \left\langle (\lambda ^{i}A^{k}g_{ik})(\lambda
^{m}A^{n}g_{mn})\right\rangle \right\vert.
\end{equation}

Here, the random-variable distribution function may be considered uniform,
with  photon polarization varying from $0$ to $\pi $:

\begin{center}
$\frac{1}{\pi }\int\limits_{0}^{\pi }\rho (\phi )d\phi =1.$
\end{center}

According to the definition,

\begin{center}
$\cos \phi =\frac{\lambda ^{i}A^{k}g_{ik}}{\sqrt{\lambda ^{i}\lambda _{i}}%
\sqrt{A^{k}A_{k}}},$

cos($\phi +\theta )=\frac{\lambda ^{n}B^{n}g_{mn}}{\sqrt{\lambda ^{m}\lambda
_{m}}\sqrt{B^{m}B_{n}}}.$
\end{center}

Hence,  correlation factor is

\begin{center}
$\left\vert M\right\vert =\left\vert \frac{1}{\pi }\int\limits_{0}^{\pi
}\rho (\phi )\cos \phi cos(\phi +\theta )d\phi +\frac{1}{\pi }%
\int\limits_{\pi }^{2\pi }\rho (\phi )\cos \phi cos(\phi +\theta )d\phi
\right\vert =\left\vert \cos \theta \right\vert .$
\end{center}

 Bell's observable  differs in our case from that calculated by Bell and
does not contradict  experiment data. Bell's inequality is not violated
in either classical or quantum cases of accounting for random fields, forces,
and waves.

\section{Conclusions}

Contemporary physics, both classical and quantum, requires a notion of
inertial reference frames. However,  to find a physical inertial frame in
a reality where there always exist random weak forces, we~suggest a
description of the motion in noninertial reference frames by means of
inclusion of higher time derivatives. They may play the role of nonlocal
hidden variables in a more general description, and can be named noninertial
mechanics, complementing both classical and quantum mechanics.
In  inertial reference frames, the derivatives of Lagrange function $L=L(q,\dot{q})$ with respect to time, coordinate, and angle are not equal to zero. This means a violation of conservation laws and  symmetries in noninertial reference frames. To preserve the symmetries in noninertial reference frames, one may use generalized Lagrange function $L=L(q,\dot{q},\ddot{q},...,q^{(n)},...)$.  Then, conservation laws are satisfied for energy, momentum, and angular momentum that not only  depend on  coordinates and velocity, but also on acceleration and high time derivatives of coordinates. 
The dynamics of mechanical systems in noninertial reference frames can be described by differential equations above the second. In this particular case, for example, when using a third-order differential equation by the method of variable replacement, it can be represented by two second-order equations. In the general case, noninertial dynamics can be described by high order differential equations.
 From the principle of equivalence, it follows that the gravitational force also has to be represented as a series. The corresponding metric is called the dark metric. The dark metric describes gravitational interaction with additional terms that lead to the description of observable effects of dark matter and dark energy. This means that the correct calculation using the dark metric leads to an abandonment of  notions of dark matter and dark energy. 
 The correctness of the results presented in this work was confirmed by comparative analysis of the use of the results of  mechanics of high-order derivatives and  experiment results of Bell's observables.

\end{document}